\documentclass[aps,prl,twocolumn,amsmath,amsfonts,amssymb,floatfix,superscriptaddress]{revtex4-2}
\usepackage{epsfig,subfigure}
\usepackage{bm}
\usepackage{color,soul,xcolor}
\usepackage{hyperref}
\usepackage{wrapfig}
\usepackage{lipsum}
\hypersetup{colorlinks=false,linkcolor=black}
\usepackage[left]{lineno}

\begin{document}
\title{Kitaev physics in the two-dimensional magnet NiPSe$_3$}

\author{Cheng Peng}
\thanks{These authors contributed equally to this work}
\email{cpeng18@stanford.edu}
\affiliation{Stanford Institute for Materials and Energy Sciences, SLAC National Accelerator Laboratory, Menlo Park, CA, USA}

\author{Sougata Mardanya}
\thanks{These authors contributed equally to this work}
\affiliation{Department of Physics and Astronomy, Howard University, Washington DC, USA}

\author{Alexander N. Petsch}
\affiliation{Stanford Institute for Materials and Energy Sciences, SLAC National Accelerator Laboratory, Menlo Park, CA, USA}
\affiliation{Linac Coherent Light Source, SLAC National Accelerator Laboratory, Menlo Park, CA, USA}

\author{Vineet Kumar Sharma}
\affiliation{Department of Physics and Astronomy, Howard University, Washington DC, USA}

\author{Shuyi Li}
\affiliation{Department of Physics, University of Florida, Gainesville, Florida 32611, USA}

\author{Chunjing Jia}
\affiliation{Department of Physics, University of Florida, Gainesville, Florida 32611, USA}

\author{Arun Bansil}
\affiliation{Department of Physics, Northeastern University, Boston, Massachusetts 02115, USA}

\author{Sugata Chowdhury}
\affiliation{Department of Physics and Astronomy, Howard University, Washington DC, USA}

\author{Joshua J. Turner}
\email{joshuat@slac.stanford.edu}
\affiliation{Stanford Institute for Materials and Energy Sciences, SLAC National Accelerator Laboratory, Menlo Park, CA, USA}

\begin{abstract}
The Kitaev interaction, found in candidate materials such as $\alpha$-RuCl$_3$, occurs through the metal~($M$)--ligand~($X$)--metal~($M$) paths of the edge-sharing octahedra because the large spin-orbit coupling (SOC) on the metal atoms activates directional spin interactions. Here, we show that even in $3d$ transition-metal compounds, where the SOC of the metal atom is negligible, heavy ligands can induce bond-dependent Kitaev interactions. In this work, we take as an example the $3d$ transition-metal chalcogenophosphate NiPSe$_3$ and show that the key is found in the presence of a sizable SOC on the Se $p$ orbital, one which mediates the super-exchange between the nearest-neighbor Ni sites. Our study provides a pathway for engineering enhanced Kitaev interactions through the interplay of SOC strength, lattice distortions, and chemical substitutions.
\end{abstract}
\maketitle

The possibility of unconventional magnetism and exotic topological excitations continues to drive intense interest in quantum materials, especially in search of the illusive quantum spin liquid. Although there is no long-range order, the spins in this state of matter are entangled. The exactly-solvable Kitaev model~\cite{KITAEV20062} has been shown to host the quantum spin liquid ground state, which arises from the bond-dependent anisotropic spin exchange in a honeycomb lattice and the associated magnetic frustration. The phase diagram~\cite{KITAEV20062} of the Kitaev model features both gapped and gapless spin-liquids, depending on the relative strengths of the various coupling parameters. While the original Kiteav model is for spin-$1/2$, numerical studies suggest that higher spins might also support the existence of quantum spin liquid states~\cite{Ks1JPSJ,Kitaevspin1PRL,PhysRevB.105.L060403,PhysRevB.102.121102}.

It has been a long journey searching for real candidate materials which realize this solvable theoretical model~\cite{doi:10.1146/annurev-conmatphys-033117-053934}. In this context, the quasi-two dimensional (2D) layered compounds, such as (Li,Na)$_2$IrO$_3$\cite{scienceSr2IrO4} and $\alpha$-RuCl$_3$~\cite{doi:10.1126/science.aah6015} have been widely investigated. Magnetic compounds with $3d$ transition metals, including CrI$_3$ and CrSiTe$_3$, also could possibly carry the Kitaev interaction even though they contain spins higher than spin-$1/2$. Recent candidate materials also include Na$_2$Co$_2$TeO$_6$ and Na$_2$Co$_2$SbO$_6$, where magnetic Co-Te-O layers are separated by a non-magnetic Na network and Co atoms with pseudospin-$1/2$ make up the layered honeycomb structure.

All the aforementioned candidate materials involve metal atoms (Ir, Ru, Co) with strong SOC in which the bond-dependent Kitaev spin interactions come into play when the nearest-neighbor electron hopping takes place through the metal~($M$)--ligand~($X$)--metal~($M$) paths of the structure through the edge-sharing octahedra~\cite{PhysRevLett.102.017205,PhysRevLett.112.077204,Kim_2022,Winter_2022}. Variations in the overlapping orbitals involved as one travels along different directions around the metal sites of the honeycomb lattice then results in anisotropic interactions, e.g. see Fig.~1 in Ref.~\cite{PhysRevResearch.3.013216} or Ref.~\cite{Kitaevspin1PRL} for instance. However, when the SOC is weak as is the case in materials with $3d$ transition metals, the open question is if Kitaev physics can still emerge? For instance, if substantial SOC effects can emerge through heavy-atom ligands, combined with Hund's coupling in the $p$ orbitals, a pathway should be possible to be created for producing bond-dependent Kitaev interactions~\cite{Kitaevspin1PRL}. It is clear that a deeper understanding in this area, and how Kitaev interactions can be activated and deactivated in magnetic materials, could provide a new basis for ``engineering'' candidate Kitaev materials.

In this Letter, we discuss the tuning of Kitaev interactions in a van der Waals family of magnets with $3d$ transition metal. For this purpose, we examine in-depth how bond-dependent anisotropic Kitaev spin interactions arise in NiPSe$_3$, and why these interactions are essentially absent in NiPS$_3$. Insight is thus gained into creating sizable Kitaev interaction terms in the Hamiltonian in Ni chalcogenophosphates, where these result when the Se $p$ orbitals experience a strong SOC effect. The spin ground state of NiPSe$_3$ is stabilized primarily through the competition between the \textit{ferromagnetic} nearest-neighbor coupling parameter $J_1$ and the \textit{antiferromagnetic} third-neighbor Heisenberg coupling $J_3$, where effects of the lattice distortion lead to off-diagonal coupling terms. An analysis of the sizes of various coupling parameters in the Hamiltonian, in relation to the SOC strength, then allows us insight into favorable chemical substitutions for enhancing Kitaev physics in these materials.

\begin{figure*}[tb]
\centering\includegraphics[width=\textwidth]{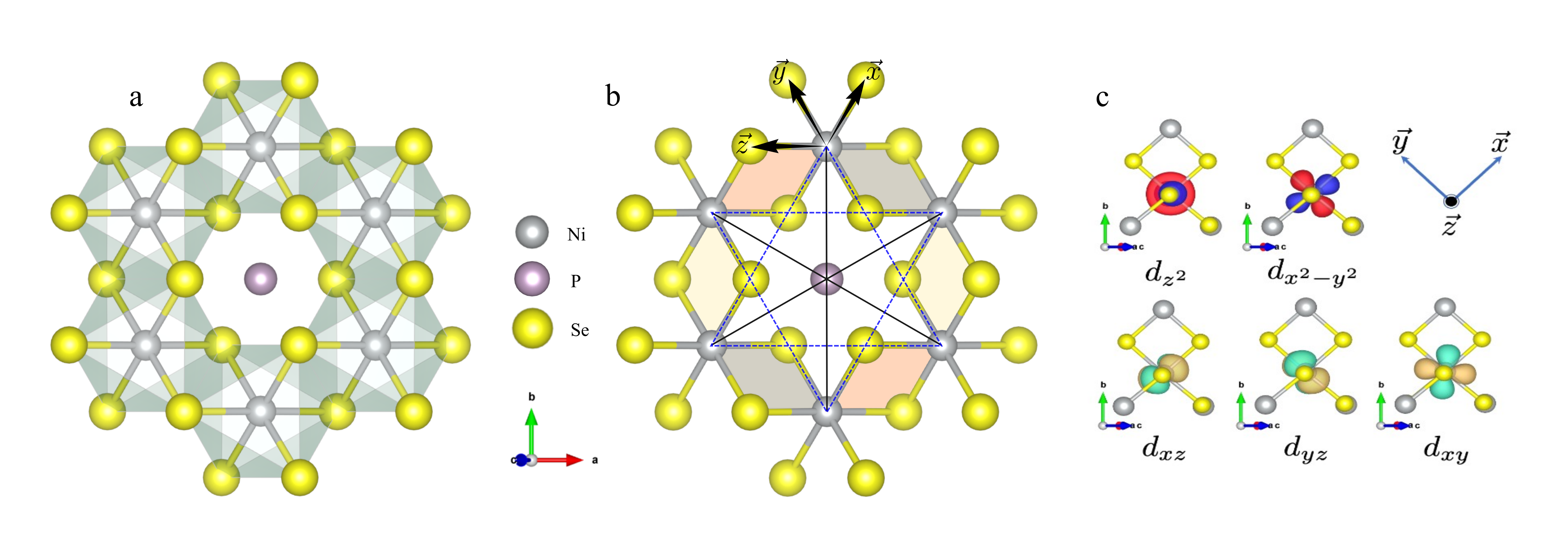}
\caption{(a) Lattice structure of a single NiPSe$_3$ layer viewed along $\mathbf{c}^*$, which is perpendicular to the $ab$-plane. Ni atoms are positioned at the centers of the octahedral cages, and the edge-sharing octahedra form the honeycomb lattice of Ni. (b) The global coordinate axes $\{\Vec{x}, \Vec{y}, \Vec{z}\}$ and the spin superexchange paths for nearest-neighbor Ni atoms are indicated by gray ($yz$-plane), orange ($zx$-plane), and yellow ($xy$-plane) markers. The second- and third-neighbor Ni atoms are shown linked with blue-dashed and black-solid lines, respectively. (c) $3d$ orbitals of Ni atoms with fully filled $t_{2g}$ orbitals in the bottom row and half-filled $e_g$ orbitals in the top row, which are aligned in accord with the global coordinate axis $\{\Vec{x}, \Vec{y}, \Vec{z}\}$.\label{fig:octahedra}}
\end{figure*}

The transition-metal chalcogenophosphates \textit{M}P\textit{X}$_3$ (TMCPs, where \textit{M} = \{Mn, Fe, Ni, Co\}, and \textit{X} = \{S, Se\}) form a family of van-der-Waals (vdW) magnetic materials~\cite{OUVRARD19851181,LEFLEM1982455}. Bulk NiP\textit{X}$_3$ (\textit{X}=\{S, Se\}) compounds, which is our focushere, have a monoclinic unit cell with space group $C2/m$ and point group $C_{2h}$. In the NiPSe$_3$ monolayer, the Ni atoms form a hexagonal structure with point group $D_{3d}$. Centers of the hexagons are occupied by phosphorus dimers (Fig.~\ref{fig:octahedra}a) and the transition-metal atoms are enclosed within octahedral cages formed by non-magnetic chalcogenide atoms that possess fully-occupied $p$ orbitals. Since NiP\textit{X}$_3$ compounds are isostructural to $\alpha$-RuCl$_3$ and CrI$_3$, their spin Hamiltonians can be constructed along similar lines if the effects of phosphorus dimers are neglected.

Due to the difficulty of growing fully Se-substituted single crystals, only a few experimental studies on NiPSe$_3$ appear in the literature~\cite{PhysRevResearch.4.023256,PhysRevMaterials.1.023402,SUN2023101188}. Therefore, construction of an effective spin model for NiPSe$_3$ also requires appeal to the spin models of other similar compounds as well as numerical simulations, although many features can be adapted from NiPS$_3$; see Supplementary Material for details~\cite{splm}. Aside from differences in the strength of the SOC, we hypothesize that the emergence of finite Kitaev spin interactions in NiPSe$_3$ is tied to the appearance of strong SOC on the ligand $p$ orbitals. We show the efficacy of our hypothesis using second-order perturbation theory.

We will neglect minor trigonal distortions of the lattice and assume an orbital splitting of the $t_{2g}$ and $e_g$ orbitals by the octahedral crystal field, as illustrated in Fig.~\ref{fig:octahedra}c. The on-site Hamiltonian for the Ni sites on a single honeycomb layer is described by the Kanamori interaction~\cite{10.1143/PTP.30.275}:
\begin{equation}
\begin{split}
  H_0 &= U_{d(p)}\sum_{a} n_{a\uparrow} n_{a\downarrow} +\frac{U'_{d(p)}}{2} \sum_{\substack{a\neq b \\ \sigma,\sigma'}}n_{a\sigma} n_{b\sigma'}\\
& -\frac{J_{H_{d(p)}}}{2}\sum_{\substack{a\neq b, \\ \sigma,\sigma'}}c^{\dagger}_{a\sigma}c^{\dagger}_{b\sigma'}c_{b\sigma}c_{a\sigma'}+J_{H_{d(p)}}\sum_{a\neq b}c^{\dagger}_{a\uparrow}c^{\dagger}_{a\downarrow}c_{b\downarrow}c_{b\uparrow}.
  \label{Eq:onsitehubbard}
\end{split}
\end{equation}

Here $U_{d(p)}$ and $U_{d(p)}^\prime$ are the intra- and inter-orbital density-density interactions, respectively, and $J_{H_{d(p)}}$ is the Hund's coupling for the spin-exchange and pair-hopping within the transition metal $e_g$ (ligand $p$) orbitals. $n_{a(b)\sigma}$ is the density operator and $c^{\dagger}_{a(b)\sigma}$($c_{a(b)\sigma}$) is the creation(annihilation) operator acting on orbital $a$($b$) and spin-$\sigma$. Here, $a$($b$) is summed over all $d$ orbitals on a transition metal site, or over all $p$ orbitals on a ligand site. Importantly, for the on-site Hamiltonian of the ligand $p$-orbitals, SOC is given by the term $H_{\text{SOC}}=\lambda_p\textbf{L}\cdot\textbf{S}$, where $\lambda_p$ denotes the SOC strength and must be taken into account.

In our case, the Hund's coupling enforces a spin-$1$ half-filling configuration of the Ni $e_g$ orbitals, while the $p$ orbitals of the ligands are completely filled and do not contribute spins in our model. Thus, only the superexchange processes between the spins of neighboring $3d^8$-Ni sites contribute to our spin-$1$ model. The $M$--$X$--$M$ superexchange is a fourth- or higher-order kinetic exchange process that involves hopping via the ligands. Note that {\it ferromagnetic} nearest-neighbor Heisenberg spin-$1$ coupling in NiP$X_3$ can only be understood through superexchange~\cite{doi:10.1021/acs.jpcc.2c00646,2020arXiv200900813C}. In contrast, direct exchange between the nearest-neighbor Ni sites is a second order kinetic exchange process that involves hopping between the transition-metal atoms without explicit involvement of the ligands, and therefore, it is {\it antiferromagnetic} and makes only a minor contribution.

The resulting spin model, if only the nearest-neighbor spin interactions are considered, is the well-known Kitaev-Heisenberg Hamiltonian:
\begin{equation}
\begin{split}
\mathcal{H} = & \sum_{\langle i,j\rangle} K^{\gamma} S^\gamma_i S^\gamma_j + J_{1}~\mathbf{S}_i\cdot \mathbf{S}_j + \Gamma \left(S^\alpha_i S^\beta_j + S^\beta_i S^\alpha_j\right)\\
& + \sum_{\langle i,j\rangle} \Gamma^\prime \left(S^\alpha_i S^\gamma_j + S^\gamma_i S^\alpha_j+S^\beta_i S^\gamma_j + S^\gamma_i S^\beta_j\right).\label{Eq:kitaevheisenberg}
\end{split}
\end{equation}

In Fig.~\ref{fig:octahedra}b, the rotations $\{\alpha, \beta, \gamma\}$ are represented by $\{y, z, x\}$ (gray), $\{z, x, y\}$ (orange) and $\{x, y, z\}$ (yellow), respectively. The symmetric off-diagonal terms $ \Gamma $ and $ \Gamma^\prime $ emerge from octahedral distortion effects~\cite{PhysRevResearch.3.013216}.

We now discuss how electron hopping along the $M$--$X$--$M$ path can be captured within the tight-binding formulation. By treating the tight-binding Hamiltonian as a perturbation to the on-site Hamiltonian, we can determine the coupling constants in the spin-$1$ model on the honeycomb lattice driven by superexchange processes. Building on the perturbation theory analysis from previous studies of NiI$_2$~\cite{Kitaevspin1PRL}, CrI$_3$~\cite{PhysRevResearch.3.013216,CrI3npj}, and NiPS$_3$~\cite{doi:10.1021/acs.jpcc.2c00646,CrI3npj}, we can then analyze NiPSe$_3$. We produce the full derivation in the Supplementary Material~\cite{splm}, and outline the main results here. The resulting Kitaev coupling strength $K^{z}$ associated, for example, with the yellow diamond in Fig.~\ref{fig:octahedra}b can be simplified as,
\begin{equation}
K^z\approx\frac{3}{2}\frac{t^4_{pd\sigma}\lambda_p^2}{(2\Delta_{pd}^2-\lambda_p\Delta_{pd}-\lambda_p^2)^2(U_d-J_{H_d})},
\end{equation}
where $\Delta_{pd}$ is the charge transfer gap between the Ni $d$ and Se $p$ orbitals. $t_{pd\sigma}$ results from the Slater-Koster formulation~\cite{PhysRev.94.1498} of the hopping integral between Ni $3d_{z^2}$ and Se $p_{z}$ orbitals. Following suggestions for NiI$_2$~\cite{Kitaevspin1PRL} and CrI$_3$~\cite{PhysRevResearch.3.013216,CrI3npj} in the literature, we assume that $K^{\gamma}$ is scaled by $\lambda_p^2$ to lowest order. Note that in the ideal case, where a perfect cubic symmetry is preserved and the hopping integrals are accurately described by Slater-Koster parameters~\cite{PhysRev.94.1498}, the off-diagonal terms in Eq.~\ref{Eq:kitaevheisenberg} will vanish. We leave them here for realistic results that can eventually be matched to experiments.

Our main goal is to answer the question on whether the Kitaev spin interaction in NiPSe$_3$ emerges simply from the replacement of S in NiPS$_3$ by Se, i.e.~to what extent can we view NiPSe$_3$ to be a NiPS$_3$-like system with stronger SOC residing on the ligand sites? An understanding of the mechanism responsible for producing Kitaev interactions could give insight into other novel effects in similar materials, such as topological magnon-phonon hybridization in FePSe$_3$~\cite{shuyi2023}. We should keep in mind, however, that the strength of the Kitaev term $K^{\gamma}$, as well as the values of the Heisenberg coupling constants and off-diagonal terms, will likely deviate significantly from our perturbation-theory based analysis here when the lattice is distorted from the perfect octahedral structure. In fact, the Ni atoms sit in a local crystal-field with environment possessing D$_3$ symmetry, which already represents a slight deviation from the cubic symmetry O$_h$. Therefore, confirmation via numerical simulations using more realistic model parameters is still recommended.

We follow the approach of Ref.~\cite{CrI3npj} based on edge-sharing octahedra to extract the exchange Hamiltonian, which is a $3\times3$ matrix encompassing exchange couplings. We obtained fitted values of $J_{\alpha}$, $J_{\beta}$ and $J_{\gamma}$ using the global axes defined in Fig.~\ref{fig:octahedra}b. Notations of Ref.~\cite{CrI3npj} is used for ease of comparison. Table~\ref{tab:DFT} compares the $K^{\gamma}$ term for NiPS$_3$ and NiPSe$_3$. Since S has negligible SOC in its $p$ orbitals, $K^{\gamma}$ for NiPS$_3$ is also negligible. However, when S is replaced with Se and the SOC in the Se $p$ orbitals is activated, $K^{\gamma}$ becomes finite. To explore the importance of the ligand SOC in the $K^{\gamma}$-term, we artificially increased the SOC strength in the ligand (third row of Table~\ref{tab:DFT}, marked with \textbf{*}). The strength of $K^{\gamma}$ is seen to increase in proportion to that of $\lambda_p^2$, consistent with our perturbation theory analysis, see Supplementary Material~\cite{splm} for details. Notably, our conclusions here are in line with those of previous studies on NiI$_2$~\cite{Kitaevspin1PRL} and CrI$_3$~\cite{PhysRevResearch.3.013216,CrI3npj}.
\begin{table}[ht]
\centering
\resizebox{\columnwidth}{!}{
\begin{tabular}{c  c  c  c  c  c  c}
\hline
 &  $J_{xx}$  &  $J_{yy}$  & $J_{zz}$  & $J_{xy}$  & $J_{yz}$ & $J_{xz}$ \\
 \hline
NiPS$_{3}$ &  $2.4822$  &  $2.4822$  &  $2.4832$  &  $0.0000$  &  $0.0000$ & $0.0000$ \\
NiPSe$_{3}$ &  $1.2800$  &  $1.2400$   &  $1.3000$  &  $0.0180$  &  $0.0040$  &  $-0.0020$\\
\textbf{*}NiPSe$_{3}$ &  $1.3988$  &  $1.2328$   &  $1.4708$  &  $0.0720$  &   $0.0120$  & $-0.0080$\\
\hline
\end{tabular}}
\resizebox{\columnwidth}{!}{
\begin{tabular}{c  c  c  c  c  c  c}
\hline
 &  $J_{\alpha}$  &  $J_{\beta}$  & $J_{\gamma}$ & $K^{\gamma}$ & $J_1$  & $J_3$  \\
 \hline
NiPS$_{3}$ &  $2.4822$  & $2.4822$  &  $2.4832$ &  $-0.001$ &  $2.4852$   & $-9.4798$  \\
NiPSe$_{3}$ &  $1.3003$  &  $1.2869$   &  $1.2328$ &  $0.0608$ &  $1.3040$   & $-12.766$  \\
\textbf{*}NiPSe$_{3}$ &  $1.4718$  &  $1.4254$   &  $1.2052$ &  $0.2434$ &  $1.4928$   & $-12.727$  \\
\hline
\end{tabular}}
\caption{\label{tab:DFT} Matrix components of our exchange Hamiltonian (in meV) featuring an effective spin-$1$ model, along with the nearest- and third-nearest-neighbor Heisenberg interactions $J_1$, $J_3$, and the bond-dependent Kitaev interaction parameter $K^{\gamma}$ for the Ni sites within a single hexagonal layer. List of all Heisenberg couplings including the second-nearest-neighbor and interlayer interactions $J_2$, $J_4$ are given in the Supplementary Material\cite{splm}. Last row (marked with \textbf{*}) gives results for NiPSe$_3$, where the SOC strength on the ligand Se is increased artificially.}
\end{table}

We emphasize here that the enhancement of $K^{\gamma}$ in this conventional Mott insulator NiPSe$_3$ occurs because the Ni atoms form a hexagonal lattice with edge-sharing octahedra and the intervening ligand Se furthermore possesses strong SOC. This result suggests that the scaling of $K^{\gamma}$ with $\lambda_p^2$ could be used more generally to control Kitaev interactions through chemical substitution of ligands in existing quantum magnets with similar lattice structures and superexchange interactions. By integrating this approach with machine learning models of structure predictions~\cite{doi:10.1126/sciadv.abn4117,npjMLstruc,Glawe_2016,doi:10.1021/ic102031h,2023arXiv231117916Z,2023arXiv231109235Y,2023arXiv231203687Z,2021arXiv211006197X,2023arXiv231214485L}, it should be possible to significantly broaden the range of candidate Kitaev materials.

Interestingly, the strongest exchange term in Table~\ref{tab:DFT} is seen to be the \textit{antiferromagnetic} third-neighbor $J_3$, which increases with Se substitution (see Supplementary Material~\cite{splm} for the full list of Heisenberg couplings). This observation is in line with a previous experimental study that shows an increase in the Néel temperature by Se substitution in NiP$X_3$~\cite{PhysRevResearch.4.023256,SUN2023101188}. An \textit{antiferromagnetic} $J_3$, which dominates in NiPSe$_3$, however, is not conducive to realizing the spin liquid state, which requires a dominant $K^{\gamma}$ term along with much smaller Heisenberg interactions~\cite{TREBST20221}. The perspective~\cite{Winter_2022} from perturbation theory shows that the energy scale of $J_3 \sim t_{pd\sigma}^4t_{pp\sigma}^2/\Delta_{pd}^4(U_d-J_{H_d})$ is substantial because $t_{pp\sigma}$ hopping integral between Se $p$ orbitals is sizable (see Supplementary Material~\cite{splm} for details on the hopping integrals). To suppress the long-range $J_3$ and obtain dominant nearest-neighbor couplings, it would be interesting to explore effects of chemical pressure and strain~\cite{PhysRevB.87.014418, doi:10.1021/acs.jpcc.2c00646} for tuning atomic distances to reduce the size of the hopping integral $t_{pp\sigma}$.

\section{Author contributions}
C.P. and J.J.T. instigated and designed this project. C.P. performed perturbation theory derivations, S.M. performed DFT+TB2J computations, A.P. provided experimental advise, V.K.S. performed DFT computations for various dopings, S.L. provided analytical advise, C.J., A.B., S.C., and J.J.T. supervised this project. C.P., A.P., S.M., and J.J.T. wrote the paper. All authors contributed to discussions and polishing of the manuscript.

\section{Acknowledgment}
This work is supported by the U.S. Department of Energy, Office of Science, Basic Energy Sciences under Award No.\@ DE-SC0022216. This research used resources of the National Energy Research Scientific Computing Center, a DOE Office of Science User Facility supported by the Office of Science of the U.S. Department of Energy under Contract No. DE-AC02-05CH11231. This research at Howard University used Accelerate ACCESS PHYS220127 and PHYS2100073. We thank Adrian Feiguin, Hong-Chen Jiang, Fangze Liu and Zhantao Chen for insightful discussions.

\begin{thebibliography}{100}

\bibitem{KITAEV20062}
Alexei Kitaev.
\newblock Anyons in an exactly solved model and beyond.
\newblock {\em Annals of Physics}, 321(1):2--111, 2006.

\bibitem{Ks1JPSJ}
Akihisa Koga, Hiroyuki Tomishige, and Joji Nasu.
\newblock {Ground-state and Thermodynamic Properties of an $S = 1$ Kitaev
  Model}.
\newblock {\em Journal of the Physical Society of Japan}, 87(6):063703, 2018.

\bibitem{Kitaevspin1PRL}
P.~Peter Stavropoulos, D.~Pereira, and Hae-Young Kee.
\newblock {Microscopic Mechanism for a Higher-Spin Kitaev Model}.
\newblock {\em Physical Review Letters}, 123:037203, 2019.

\bibitem{PhysRevB.105.L060403}
Yu-Hsueh Chen, Jozef Genzor, Yong~Baek Kim, and Ying-Jer Kao.
\newblock {Excitation spectrum of spin-1 Kitaev spin liquids}.
\newblock {\em Physical Review B}, 105:L060403, 2022.

\bibitem{PhysRevB.102.121102}
Xiao-Yu Dong and D.~N. Sheng.
\newblock {Spin-1 Kitaev-Heisenberg model on a honeycomb lattice}.
\newblock {\em Physical Review B}, 102:121102, 2020.

\bibitem{doi:10.1146/annurev-conmatphys-033117-053934}
M.~Hermanns, I.~Kimchi, and J.~Knolle.
\newblock {Physics of the Kitaev Model: Fractionalization, Dynamic
  Correlations, and Material Connections}.
\newblock {\em Annual Review of Condensed Matter Physics}, 9(1):17--33, 2018.

\bibitem{scienceSr2IrO4}
B.~J. Kim, H.~Ohsumi, T.~Komesu, S.~Sakai, T.~Morita, H.~Takagi, and T.~Arima.
\newblock {Phase-Sensitive Observation of a Spin-Orbital Mott State in
  Sr$_2$IrO$_4$}.
\newblock {\em Science}, 323(5919):1329--1332, 2009.

\bibitem{doi:10.1126/science.aah6015}
Arnab Banerjee, Jiaqiang Yan, Johannes Knolle, Craig~A. Bridges, Matthew~B.
  Stone, Mark~D. Lumsden, David~G. Mandrus, David~A. Tennant, Roderich
  Moessner, and Stephen~E. Nagler.
\newblock {Neutron scattering in the proximate quantum spin liquid
  $\alpha$-RuCl$_3$}.
\newblock {\em Science}, 356(6342):1055--1059, 2017.

\bibitem{PhysRevLett.102.017205}
G.~Jackeli and G.~Khaliullin.
\newblock {Mott Insulators in the Strong Spin-Orbit Coupling Limit: From
  Heisenberg to a Quantum Compass and Kitaev Models}.
\newblock {\em Physical Review Letters}, 102:017205, 2009.

\bibitem{PhysRevLett.112.077204}
Jeffrey~G. Rau, Eric Kin-Ho Lee, and Hae-Young Kee.
\newblock {Generic Spin Model for the Honeycomb Iridates beyond the Kitaev
  Limit}.
\newblock {\em Physical Review Letters}, 112:077204, 2014.

\bibitem{Kim_2022}
Chaebin Kim, Heung-Sik Kim, and Je-Geun Park.
\newblock {Spin-orbital entangled state and realization of Kitaev physics in
  $3d$ cobalt compounds: a progress report}.
\newblock {\em Journal of Physics: Condensed Matter}, 34(2):023001, 2021.

\bibitem{Winter_2022}
Stephen~M Winter.
\newblock Magnetic couplings in edge-sharing high-spin $d^7$ compounds.
\newblock {\em Journal of Physics: Materials}, 5(4):045003, 2022.

\bibitem{PhysRevResearch.3.013216}
P.~Peter Stavropoulos, Xiaoyu Liu, and Hae-Young Kee.
\newblock {Magnetic anisotropy in spin-3/2 with heavy ligand in honeycomb Mott
  insulators: Application to ${\mathrm{CrI}}_{3}$}.
\newblock {\em Physical Review Research}, 3:013216, 2021.

\bibitem{OUVRARD19851181}
G.~Ouvrard, R.~Brec, and J.~Rouxel.
\newblock {Structural determination of some MPS$_3$ layered phases (M = Mn, Fe,
  Co, Ni and Cd)}.
\newblock {\em Materials Research Bulletin}, 20(10):1181--1189, 1985.

\bibitem{LEFLEM1982455}
G.~{Le Flem}, R.~Brec, G.~Ouvard, A.~Louisy, and P.~Segransan.
\newblock {Magnetic interactions in the layer compounds $M$P$X_3$ ($M$ = Mn,
  Fe, Ni; $X$ = S, Se)}.
\newblock {\em Journal of Physics and Chemistry of Solids}, 43(5):455--461,
  1982.

\bibitem{PhysRevResearch.4.023256}
Rabindra Basnet, Kamila~M. Kotur, Milosz Rybak, Cory Stephenson, Samuel Bishop,
  Carmine Autieri, Magdalena Birowska, and Jin Hu.
\newblock {Controlling magnetic exchange and anisotropy by nonmagnetic ligand
  substitution in layered $M\mathrm{P}{X}_{3}$ ($M=\mathrm{Ni}$, Mn;
  $X=\mathrm{S}$, Se)}.
\newblock {\em Physical Review Research}, 4:023256, 2022.

\bibitem{PhysRevMaterials.1.023402}
J.-Q. Yan, B.~C. Sales, M.~A. Susner, and M.~A. McGuire.
\newblock {Flux growth in a horizontal configuration: An analog to vapor
  transport growth}.
\newblock {\em Physical Review Materials}, 1:023402, 2017.

\bibitem{SUN2023101188}
Hualei Sun, Liang Qiu, Yifeng Han, Enkui Yi, Junlong Li, Mengwu Huo, Chaoxin
  Huang, Hui Liu, Manrong Li, Weiliang Wang, Dao-Xin Yao, Benjamin~A. Frandsen,
  Bing Shen, Yusheng Hou, and Meng Wang.
\newblock {Coexistence of zigzag antiferromagnetic order and superconductivity
  in compressed NiPSe$_3$}.
\newblock {\em Materials Today Physics}, 36:101188, 2023.

\bibitem{splm}
See the Supplemental material at [url] for complementrary results, more
  analitical and numerical details.

\bibitem{10.1143/PTP.30.275}
Junjiro Kanamori.
\newblock {Electron Correlation and Ferromagnetism of Transition Metals}.
\newblock {\em Progress of Theoretical Physics}, 30(3):275--289, 1963.

\bibitem{doi:10.1021/acs.jpcc.2c00646}
Carmine Autieri, Giuseppe Cuono, Canio Noce, Milosz Rybak, Kamila~M. Kotur,
  Cliò~Efthimia Agrapidis, Krzysztof Wohlfeld, and Magdalena Birowska.
\newblock {Limited Ferromagnetic Interactions in Monolayers of MPS$_3$ (M = Mn
  and Ni)}.
\newblock {\em The Journal of Physical Chemistry C}, 126(15):6791--6802, 2022.

\bibitem{2020arXiv200900813C}
Swati {Chaudhary}, Alon {Ron}, David {Hsieh}, and Gil {Refael}.
\newblock {Controlling ligand-mediated exchange interactions in periodically
  driven magnetic materials}.
\newblock {\em arXiv e-prints}, page arXiv:2009.00813, 2020.

\bibitem{CrI3npj}
Changsong Xu, Junsheng Feng, Hongjun Xiang, and Laurent Bellaiche1.
\newblock {Interplay between Kitaev interaction and single ion anisotropy in
  ferromagnetic CrI$_3$ and CrGeTe$_3$ monolayers}.
\newblock {\em npj Computational Materials}, 4, 2018.

\bibitem{PhysRev.94.1498}
J.~C. Slater and G.~F. Koster.
\newblock {Simplified LCAO Method for the Periodic Potential Problem}.
\newblock {\em Physical Review}, 94:1498--1524, 1954.

\bibitem{shuyi2023}
Jiaming Luo, Shuyi Li, Zhipeng Ye, Rui Xu, Han Yan, Junjie Zhang, Gaihua Ye,
  Lebing Chen, Ding Hu, Xiaokun Teng, William~A. Smith, Boris~I. Yakobson,
  Pengcheng Dai, Andriy~H. Nevidomskyy, Rui He, and Hanyu Zhu.
\newblock {Evidence for Topological Magnon–Phonon Hybridization in a 2D
  Antiferromagnet down to the Monolayer Limit}.
\newblock {\em Nano Letters}, 23(5):2023--2030, 2023.

\bibitem{doi:10.1126/sciadv.abn4117}
Rhys E.~A. Goodall, Abhijith~S. Parackal, Felix~A. Faber, Rickard Armiento, and
  Alpha~A. Lee.
\newblock Rapid discovery of stable materials by coordinate-free coarse
  graining.
\newblock {\em Science Advances}, 8(30):eabn4117, 2022.

\bibitem{npjMLstruc}
HC. Wang, S.~Botti, and M.A.L. Marques.
\newblock Predicting stable crystalline compounds using chemical similarity.
\newblock {\em npj Computational Materials}, 7:12, 2021.

\bibitem{Glawe_2016}
Henning Glawe, Antonio Sanna, U.~Gross E.\, K.\, and Miguel A.~L. Marques.
\newblock {The optimal one dimensional periodic table: a modified Pettifor
  chemical scale from data mining}.
\newblock {\em New Journal of Physics}, 18(9):093011, 2016.

\bibitem{doi:10.1021/ic102031h}
Geoffroy Hautier, Chris Fischer, Virginie Ehrlacher, Anubhav Jain, and Gerbrand
  Ceder.
\newblock {Data Mined Ionic Substitutions for the Discovery of New Compounds}.
\newblock {\em Inorganic Chemistry}, 50(2):656--663, 2011.
\newblock PMID: 21142147.

\bibitem{2023arXiv231117916Z}
Ruiming {Zhu}, Wei {Nong}, Shuya {Yamazaki}, and Kedar {Hippalgaonkar}.
\newblock {WyCryst: Wyckoff Inorganic Crystal Generator Framework}.
\newblock {\em arXiv e-prints}, page arXiv:2311.17916, 2023.

\bibitem{2023arXiv231109235Y}
Mengjiao {Yang}, KwangHwan {Cho}, Amil {Merchant}, Pieter {Abbeel}, Dale
  {Schuurmans}, Igor {Mordatch}, and Ekin {Dogus Cubuk}.
\newblock {Scalable Diffusion for Materials Generation}.
\newblock {\em arXiv e-prints}, page arXiv:2311.09235, 2023.

\bibitem{2023arXiv231203687Z}
Claudio {Zeni}, Robert {Pinsler}, Daniel {Z{\"u}gner}, Andrew {Fowler}, Matthew
  {Horton}, Xiang {Fu}, Sasha {Shysheya}, Jonathan {Crabb{\'e}}, Lixin {Sun},
  Jake {Smith}, Bichlien {Nguyen}, Hannes {Schulz}, Sarah {Lewis}, Chin-Wei
  {Huang}, Ziheng {Lu}, Yichi {Zhou}, Han {Yang}, Hongxia {Hao}, Jielan {Li},
  Ryota {Tomioka}, and Tian {Xie}.
\newblock {MatterGen: a generative model for inorganic materials design}.
\newblock {\em arXiv e-prints}, page arXiv:2312.03687, 2023.

\bibitem{2021arXiv211006197X}
Tian {Xie}, Xiang {Fu}, Octavian-Eugen {Ganea}, Regina {Barzilay}, and Tommi
  {Jaakkola}.
\newblock {Crystal Diffusion Variational Autoencoder for Periodic Material
  Generation}.
\newblock {\em arXiv e-prints}, page arXiv:2110.06197, 2021.

\bibitem{2023arXiv231214485L}
Fangze {Liu}, Zhaotao {Chen}, Tianyi {Liu}, Yu~{Lin}, Joshua~J. {Turner}, and
  Chunjing {Jia}.
\newblock {Self-Supervised Generative Models for Crystal Structures}.
\newblock {\em arXiv e-prints}, page arXiv:2312.14485, 2023.

\bibitem{TREBST20221}
Simon Trebst and Ciarán Hickey.
\newblock Kitaev materials.
\newblock {\em Physics Reports}, 950:1--37, 2022.

\bibitem{PhysRevB.87.014418}
A.~V. Ushakov, D.~A. Kukusta, A.~N. Yaresko, and D.~I. Khomskii.
\newblock {Magnetism of layered chromium sulfides $M$CrS$_{2}$
  ($M=\mathrm{Li}$, Na, K, Ag, and Au): A first-principles study}.
\newblock {\em Physical Review B}, 87:014418, 2013.

\end{thebibliography}

\end{document}